\documentclass[%
reprint,
nofootinbib,
 amsmath,amssymb,
 aps,
floatfix,
showkeys
]{revtex4-2}

\usepackage{graphicx}
\usepackage{dcolumn}
\usepackage{bm}
\usepackage{setspace}
\usepackage{amsmath}
\usepackage{tabularx}
\usepackage{color}

\newcommand{\overbar}[1]{\mkern 1.5 mu\overline{\mkern-1.5mu#1\mkern-1.5mu}\mkern 1.5mu}
\newcommand{\orcid}[1]{\href{https://orcid.org/#1}{\textcolor[HTML]{A6CE39}{\aiOrcid}}}

\begin{document}
\preprint{APS/123-QED}

\title{Modeling the Past Hypothesis: A Mechanical Cosmology}

\author{Jordan Scharnhorst$^{1,2,*}$}
\author{Anthony Aguirre$^{1,2}$}
\affiliation{%
Physics Department, University of California, Santa Cruz
}%
\affiliation{Santa Cruz Institute for Particle Physics}
\email{Corresponding author: jscharnh@ucsc.edu}
\date{\today}

\begin{abstract}
There is a paradox in the standard model of cosmology. How can matter in the early universe have been in thermal equilibrium, indicating maximum entropy, but the initial state also have been low entropy (the ``past hypothesis"), so as to underpin the second law of thermodynamics? The problem has been highly contested, with the only consensus being that gravity plays a role in the story, but with the exact mechanism undecided. In this paper, we construct a well-defined mechanical model to study this paradox. We show how it reproduces the salient features of standard big-bang cosmology with surprising success, and we use it to produce novel results on the statistical mechanics of a gas in an expanding universe. We conclude with a discussion of potential uses of the model, including the explicit computation of the time-dependent coarse-grained entropies needed to investigate the past hypothesis.

\end{abstract}


\maketitle

\section{Introduction}

In contrast to the existence of ubiquitous time-asymmetric phenomena, known microscopic physical laws are all time-reversal (or CPT) invariant. It has been argued that the boundary conditions of the theory can then explain the asymmetry. This is the foundation of the past hypothesis -- the hypothesis that the early universe, or initial state of the universe, had a very low entropy compared to the entropy today. The past hypothesis is widely presumed to account for the arrow of time and the irreversible phenomena we observe in thermodynamic systems \cite{sep-time-thermo, wallace2009gravity}.

It is also known from measurements of the cosmic microwave background and basic theoretical consistency that the constituents of the early universe were in thermal equilibrium. However, this combines with the past hypothesis to produce a paradox, as equilibrium states have maximum entropy by definition. Competing resolutions have been proposed in the literature, with two main ideas emerging \cite{Penrose:1980ge,Rovelli:2019}: 
\begin{enumerate}
    \item[1.] The clustering explanation;
    \item[2.] The expansion explanation.
\end{enumerate}

These resolutions argue that the early universe was actually quite {\em out} of equilibrium when taking gravity into account. The first argument holds that a uniform matter distribution is actually \textit{lower entropy} than a clumped distribution, since matter clumps under the influence of Newtonian gravity. The second argument holds that low entropy, in the non-equilibrium sense, was due to cosmological expansion.  This argument makes the point that expansion changes the equilibrium state and does so faster than the matter can attain its equilibrium. The early smallness of the scale factor acts as a constraint that leaves the matter degrees of freedom stuck in a state that has lower entropy than ones in which the constraint is removed. Rovelli \cite{Rovelli:2019} argues that the cosmology explanation can be seen explicitly in a suitable model calculation.

Earman~\cite{Earman:2006dgv} has critiqued this competition and argues that ``a resolution of the controversy is not to be obtained by means of intuition pumps but rather through precise model calculations." How might Earman's vision be realized?

Such a model would first and foremost require a well-defined state space since one needs such a state space to rigorously discuss entropy and entropy increase. This space is usually a symplectic manifold of coordinates and momenta or a Hilbert space. With this space, a coarse-graining can be defined, which partitions the space into macrostates that are collections of microstates.~\footnote{An alternative approach \cite{Brand_o_2008,gemmer2009quantum} to entropy increase in a system involves tracing or marginalizing over degrees of freedom outside the system, but in the case of our model cosmology there are none.} Beyond this, a model should truly capture the salient features of cosmology: expansion rates, equations of state, freeze-out, global geometry, dark energy, and structure formation.

Work has been done on aspects of calculating the entropy of matter during gravitational clustering \cite{Padmanabhan:2008xf} and in an expanding universe \cite{Aoki, Frautschi:1981xu, 2008PhLRv...5..225L}. Many argue that the expansion of the universe is isentropic (or nearly so), meaning that there is no net entropy change in matter due to expansion. Some argue that entropy continues to increase throughout expansion, but at a rate that is too slow to keep up with the growth of the maximum entropy \cite{Weber1988EntropyIA,Layzer1990-LAYCTG}. Approaches to kinetic theory in an expanding universe based on the Boltzmann equation have been explored \cite{Kremer_2002,Kremer_2014,Haba_2017,Tindall_2017,Zimdahl_1998}, but these treat spacetime and gravity separately and as background. The extent to which self-contained models have been constructed is minimal. We seek to provide an analog model for cosmology, in which the main modalities of entropy change can be studied explicitly and in a self-contained way.

In this paper, we will define the model, show how it reproduces the key aspects of big-bang cosmology, interpret standard cosmological results in this lens, show an application to non-equilibrium statistical mechanics, and end with a discussion of how it can be used to compute coarse-grained entropy, which is left for later work.

\section{Statistical Mechanics and Gravity}

The intersection of thermodynamics and gravity is a rich subject. (For a general review, see \cite{Padmanabhan:2009vy}.) Since the development of general relativity, we have learned that black holes radiate, particles can be created in a gravitational field, there is a correspondence between Anti-de Sitter space and conformal field theories, and discovered the gravitational path integral.

While these are all quantum effects, classical studies of cosmology and gravity are still well-motivated \cite{HU_2011}. Some of gravity's weird features (potentially even including the relation between area and entropy \cite{Oppenheim}) are manifest in a classical setting due to gravity's long-range and unshielded nature.

The thermodynamic nature of gravitational systems, even classical, is surely strange. It is well-known that gravitational systems have negative heat-capacities -- which is why black holes get \textit{hotter} rather than \textit{colder} as they evaporate and give off heat. Similarly, a gas with an attractive Newtonian gravitational potential has no equilibrium -- it will form a `core' that becomes increasingly hot in a runaway process, while emitting heat in the form of particles that have escaped the potential energy barrier, called a `halo' \cite{Padmanabhan:2008xf}. In the absence of cutoffs, the entropy will diverge as a function of time. ``Non-equilibrium" is the general description for the statistical physics of gravitational systems \cite{HU_2011}. Our model will not fully tame this strangeness; but it can perhaps segregate types of strangeness from each other, in particular those that stem from quantum rather than classical gravity, and those that relate to clustering rather than those that would persist even in the absence of gravitational structure growth.

\begin{figure*}[t]
\centering
    \includegraphics[scale=.7]{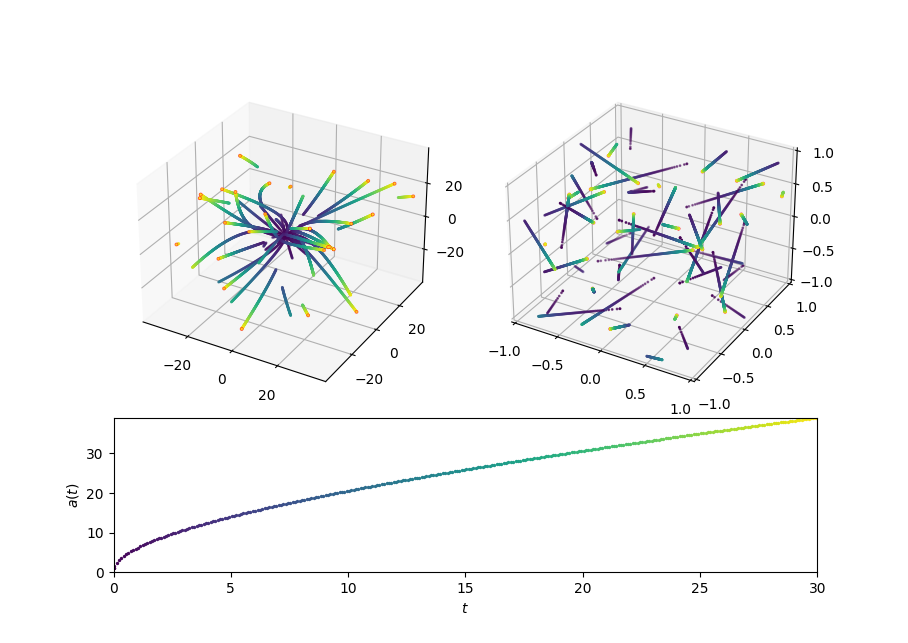}
    \caption{Left: Paths of $N=27$ non-interacting particles in physical coordinates. Partial trails are joined by \textit{comoving} periodic boundary conditions. Right: Paths of $N=27$ non-interacting particles in comoving coordinates, in which the slowing down of the comoving velocities can be seen. Bottom: The scale factor $a(t)$ with the color-mapped time dependence.}
    \label{3D}
\end{figure*}

\section{The Model}

Our model, defined by Eq. (\ref{Ham}) below, is closed, well-behaved, and consists only of a single, classical Hamiltonian with $6N+1$ dynamical degrees of freedom, where $N$ is the number of particles.
The $6N$ degrees come from the particle positions and momenta, and $2$ degrees come from $a$ and $p_a$, with one degree removed through the ``Hamiltonian constraint"  that $H=0$, which should be satisfied in theories with dynamical degrees of freedom for gravity. (See \cite{Arnowitt_2008} for a comprehensive and \cite{corichi2022introduction} for a modern review of the ADM formalism, in which the same point arises.) 

In reducing the metric degrees of freedom to just the scale factor, our model is like a Minisuperspace model. These models describe isotropic and homogenous gravitational systems in which fields couple to the FLRW scale factor $a(t)$. They then take the scale factor as a single degree of freedom and attempt canonical or path-integral quantization; this is the foundation of quantum cosmology \cite{Bojowald_2015}. The dynamics for $a$ are induced via a dependence of the Lagrangian on $\dot a$. Unlike a minisuperspace model, our model does not assume homogeneity of the matter, but it does treat the matter and scale factor degrees of freedom on the same footing.\footnote{Although the interpretation of the scale factor becomes unclear if the particle distribution is highly nonuniform, there is nothing formally wrong with the model in this regime.}

All particle coordinates $x_i$ and momenta $p_i$ refer to the comoving coordinates and comoving momenta, compared to the physical coordinates $ax_i$ and physical momenta ${p_i}/{a}.$ For convenience we set $a(0)=1,$ else the physical momenta would read ${p_i}/{(a/a(0))}.$
The Hamiltonian reads:
\begin{equation}\label{Ham}
\begin{split}
&H=\mathcal{N}\Big[\frac{1}{8\pi G}\left(-\left(8\pi G\right)^2\frac{{p_a}^2}{12a}-3ka-\Lambda a^3\right)\\
+\sum_{i}&{\sqrt{m_i^2+\frac{p_{r_i}^2 (1-kr_i^2)}{a^2}+\frac{p_{\theta_i}^2}{a^2r_i^2}+\frac{p_{\varphi_i}^2}{a^2r^2\sin^2(\theta_i)}}}\\
&\;\;\;\;\;\;\;\;\;\;\;\;\;\;\;\;+\sum_{i,j}V_{ij}(a|\boldsymbol r_i-\boldsymbol r_j|)\Big],
\end{split}
\end{equation}
where $k$ is either $\pm 1$ or $0$ and defines the global geometry of the spacetime, and $\Lambda$ is the cosmological constant. Put simply, 
$$\begin{aligned}
H\sim&\text{ Kinetic Term for }a\,+ \text{ Potential terms for }a\, \\
&+ \text{ Relativistic Particles } + \text{ Interactions}.
\end{aligned}$$

From the corresponding Lagrangian (derived in the Appendix), we can compute the conjugate momenta $p_a=-{3a\dot a}/({4\pi G \mathcal{N}})$ and $ p_{x_i} \sim {(mv_i/a^2)}/{\sqrt{1-a^2 v^2}}.$ The $p_{x_i}$ have a dependence on $k$, and the $\sim$ becomes an equality when $k=0$.

We immediately notice a few things. First, the kinetic term for $a$ is both negative and non-separable (there is a coupling between a coordinate and momentum). In minisuperspace models (which have the same Hamiltonian structure), gravitational energy has the opposite sign of the energy in matter -- which is necessary so that the Hamiltonian constraint $H=0$ can hold. There is an extraneous variable $\mathcal{N}$, called the \textit{lapse function}, used as a Lagrange multiplier to enforce this constraint, and the lack of time dependence indicates that energy is conserved. The inter-particle interactions depend on the physical distance between particles, $a|\boldsymbol r_i-\boldsymbol r_j|$, rather than the comoving distance $|\boldsymbol r_i-\boldsymbol r_j|$.

The model is a good approximation for cosmology as $N\rightarrow \infty$, assuming the spatial distributions become uniform. Inflationary physics is, in principle, simple to include as the homogenous inflaton is a single degree of freedom that can be treated mechanically. However, at the time and energy scales of inflation, classical mechanics does not apply. The model does not capture chemical effects, e.g. particle conversion, but it does capture kinetic equilibrium and decoupling, which underpin the chemical effects. The thermal and energetic effects of massless particles~\footnote{Although the model treats relativistic energies correctly, it is manifestly {\em not} a special- or general-relativistic model.} are simple to include by taking the limit as $m\rightarrow 0$.

\section{Equations of Motion}
Here and in the following, unless otherwise stated, $|\boldsymbol p_i|^2$ is understood to depend on $k$ and reduces to the ``flat" definition $|\boldsymbol p_i|^2=p_{x_i}^2+p_{y_i}^2+p_{z_i}^2$ upon setting $k=0$.

$\mathcal{N}$ is non-dynamical but treated as a coordinate for the purpose of enforcing the constraint $H=0$. Its conjugate momentum $p_N$ is 0 via $p_\mathcal{N}=\frac{\partial L}{\partial \mathcal{\dot N}}$, where $L$ is the corresponding model Lagrangian, derived in the appendix. Trivially, $\dot p_\mathcal{N} = 0$ and $\mathcal{\dot N}=0$.

\subsection{Hamilton's Equations}
Hamilton's equations read:
\begin{equation}\label{HamN}
\begin{aligned}
\dot p_\mathcal{N} = -\frac{\partial H}{\partial \mathcal{N}} = \frac{1}{8\pi G}\left(-\left(8\pi G\right)^2\frac{{p_a}^2}{12a}-3ka-\Lambda a^3\right)\\+\sum_{i}\sqrt{m_i^2+\frac{|\boldsymbol p_i|^2}{a^2}}+\sum_{i,j} V_{ij}(a|\boldsymbol r_i-\boldsymbol r_j|)=0,
\end{aligned}
\end{equation}
where the last equality follows from the fact that $p_\mathcal{N}$ is a constant.

\begin{equation}\label{HamA}
\dot a=\frac{\partial H}{\partial p_a}=\mathcal{N}\Big [-\frac{4}{3}\pi G\frac{p_a}{a}\Big ]
\end{equation}
\begin{equation}\label{HamPa}
\begin{aligned}
\dot p_a=-\frac{\partial H}{\partial a}=\mathcal{N}\Big[-\frac{2}{3}\pi G\frac{p_a^2}{a^2} + \frac{3k}{8\pi G} + 3\Lambda a^2 + \\ \frac{1}{a^3}\sum_i\frac{|\boldsymbol p_i|^2}{\sqrt{m^2+\frac{|\boldsymbol p_i|^2}{a^2}}}-\sum_{i,j} \frac{\partial V_{ij}}{\partial {(a|\boldsymbol r_i-\boldsymbol r_j|)}}{|\boldsymbol r_i-\boldsymbol r_j|}\Big ]
\end{aligned}
\end{equation}

Using Eq. (\ref{HamA}), we can write the Hubble parameter as $h=\dot a/a=-(4\pi G \mathcal{N}/3a^2)p_a$. 
\subsection{Friedmann Equations}
The well-known Friedmann equations are a set of two equations governing the dynamics of spacetime and matter for the FLRW metric. 
Any reasonable model of cosmology should reproduce the Friedman equations, or something equivalent. We will see that this is the case.
Combining Hamilton's equations for $a$ upon gauging $\mathcal{N}=1$, we have
\begin{equation}\label{Fried1}
   2\frac{\ddot a}{a}+\frac{{\dot a}^2}{a^2}+\frac{k}{a^2}=-8\pi GP+\Lambda,
\end{equation}
where 
\begin{equation}\label{P}
\begin{split}
P= \frac{1}{3}\Biggl(\frac{1}{a^3}\sum_i \frac{|\boldsymbol p_i|^2 / a^2}{\sqrt{m_i^2+\frac{|\boldsymbol p_i|^2}{a^2}}} - \\ \sum_{i,j} \frac{\partial V_{ij}}{\partial {(a|\boldsymbol r_i-\boldsymbol r_j|)}}{|\boldsymbol r_i-\boldsymbol r_j|}\Biggr),
\end{split}
\end{equation}
and $|\boldsymbol r_i-\boldsymbol r_j|$ is the comoving interparticle distance. This is a linear combination of the standard two Friedmann equations. The definition of $P$ in the non-interacting case agrees with that derived for a relativistic gas thermodynamically \cite{DE_BERREDO_PEIXOTO_2005}.

Rearranging Eq.~(\ref{HamN}) and using the relation $p_a=-3a\dot a/(4\pi G),$ we derive
\begin{equation}\label{Fried2}
    \frac{k}{a^2}+\frac{{\dot a}^2}{a^2}=\frac{8\pi G\rho + \Lambda}{3},
\end{equation}
where
\begin{equation}\label{rho}
      \rho= \frac{1}{a^3}\left(\sum_i\sqrt{m_i^2+\frac{|\boldsymbol p_i|^2}{a^2}} + \sum_{i,j}V_{ij}(a|\boldsymbol r_i-\boldsymbol r_j|)\right).
\end{equation}
The interpretation is quite clear, $\rho= E/V$, where $V$ is the physical volume and $E$ is the total energy of the particles. Together, these let us derive
\begin{equation}\label{Fried3}
\frac{\ddot a}{a}=-\frac{4\pi G}{3}\left( \rho + 3P \right)+\frac{\Lambda}{3},
\end{equation}
the other Friedmann equation. 

\begin{figure}
    \includegraphics[scale=.65]{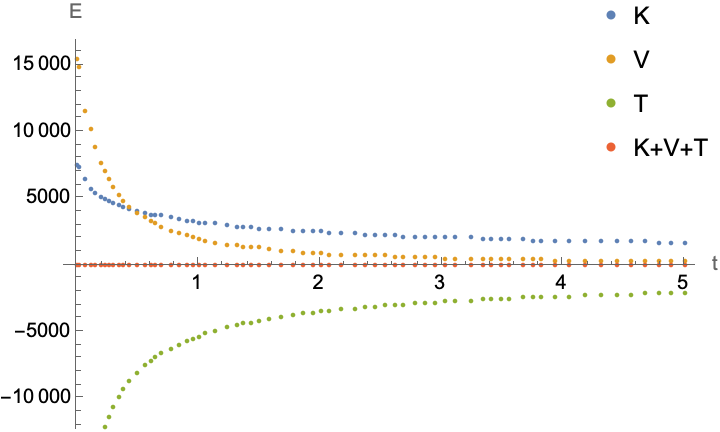}
    \caption{Example time dependence of the terms in Eq. (\ref{Ham}) for $k=\Lambda=0$. $K$ is the particle kinetic energy, $V$ is particle potential energy, $T$ is total gravitational energy. In red is the total energy, which is 0. $V(r)$ is an inverse power law.}
    \label{Energies}
\end{figure}

Eq. (\ref{Fried1}) is exactly the Euler-Lagrange equation of the model Lagrangian, and needs to be combined with the $00$ component equation of the Einstein Equations (conservation of energy), Eq. (\ref{Fried2}), to obtain Eq. (\ref{Fried3}).

\subsection{Particles}
In the absence of interactions, particles follow the standard FLRW geodesics. For $k=0$ and with interactions, they have the following Hamilton's equations:
\begin{equation}\label{velocities}
\dot {\boldsymbol x}_i=\mathcal{N}\frac{ \boldsymbol p_{i}/a^2}{\sqrt{m^2+\frac{|\boldsymbol p_i|^2}{a^2}}}
\end{equation}
and 
\begin{equation}\label{forces}
\dot {\boldsymbol p}_i = -\mathcal{N}a\sum_{j} \frac{\partial V_{ij}}{\partial {(a|\boldsymbol r _i-\boldsymbol r_j|)}}.
\end{equation}

The peculiar velocity of a particle, which enters directly into the Lagrangian (see Appendix), is $\dot {\boldsymbol x}_i a$, and the recessional velocity is $\dot a {\boldsymbol x}_i = (\dot a/a)(a{\boldsymbol x}_i).$ The total velocity then is the sum of the two, equal to the time derivative of the physical distance: $v_{tot}=\frac{d}{dt}(a{\boldsymbol x}_i)=\dot a {\boldsymbol x}_i + \dot {\boldsymbol x}_i a.$

 In FIG. \ref{3D}, it is evident that the particles slow down in comoving coordinates; for non-relativistic (NR) particles, the peculiar velocity has a $1/a$ scaling, which acts as friction. For highly-relativistic (HR) particles, the peculiar velocity is approximately constant.

So far, we have seen that the equations of motion are the Friedmann equations and the particles follow their (interacting) geodesics. The particles have built-in redshift/blueshift, as the comoving momenta $p_{i}$ couple directly to $a$, and thus energy is able to be transferred between the gravitational field and matter.

\section{Reproducing Cosmology Numerically}

Although many results can be derived directly from the model, it is useful to explore (and verify) aspects of the model numerically. For our purposes, $N<1000$ suffices. We primarily explore repulsive potentials; with an attractive potential, the calculation produces clustering, but to study this in detail would require more sophisticated (but well known) $N$-body methods.\footnote{The caveat is that the momenta also need to be treated, as they couple to $a$. The numerical price to pay is small, with a number of ``momentum" computations of order $N$, compared to the number of particle-particle interaction computations, which is of order $N^2.$}

\subsection{Simulation Methods}
To simulate an example model, we evolve Hamilton's equations Eqs.~(\ref{HamN}) - (\ref{HamPa}) and (\ref{velocities}) - (\ref{forces}) with $2N$ particles on a flat space with no cosmological constant. The constraint that $H=0$, equivalent to Eq.~(\ref{HamN}), is imposed via an initial condition on $p_a$. We impose periodic boundary conditions on the comoving coordinates, making the space topologically $\mathbb{T}^3$. Units are chosen such that $a(0)=1$ and $c = 1.$ The positions are initialized on a uniform lattice and the momenta are initialized with random directions and sampled from a Maxwell-Juttner distribution, the equilibrium distribution of a relativistic gas. (See Sec. VII)

To study thermalization, a suitable short-ranged potential must be chosen. In FLRW universes, particles interact via the \textit{physical distance} between them, not the comoving distance, so the interaction potential should depend on $a$ in addition to the comoving coordinates. The potential must be periodic, with comoving period equal to the box size, and the forces should be continuous across the boundary. Beyond these requirements, we are free to choose any potential as the thermodynamic details depend weakly on the interactions.  

In particular we choose $V\sim {1}/{(ar)^\alpha}$, with $\delta x_i\rightarrow \sin\left(\frac{\pi \delta x_i}{2}\right)$ to satisfy periodicity, so
\begin{equation}\label{vij}
     V_{ij}=\frac{q}{\left(a\sqrt{\sin(\frac{\pi \delta x}{2})^2+\sin(\frac{\pi \delta y}{2})^2+\sin(\frac{\pi \delta z}{2})^2}\right)^\alpha},
\end{equation}
where $\delta x=x_i-x_j.$ This potential is periodic in the three comoving coordinates, depends correctly on $a$, is time independent, and approaches the corresponding central potential $V\sim {1}/{(ar)^\alpha}$ rapidly near $r=0$, and is a good approximation across the entire region. See the appendix for a comparison with the corresponding central potential for $\alpha=4$.

\subsection{Numerical Cosmology}

\begin{figure}
    \includegraphics[scale=.65]{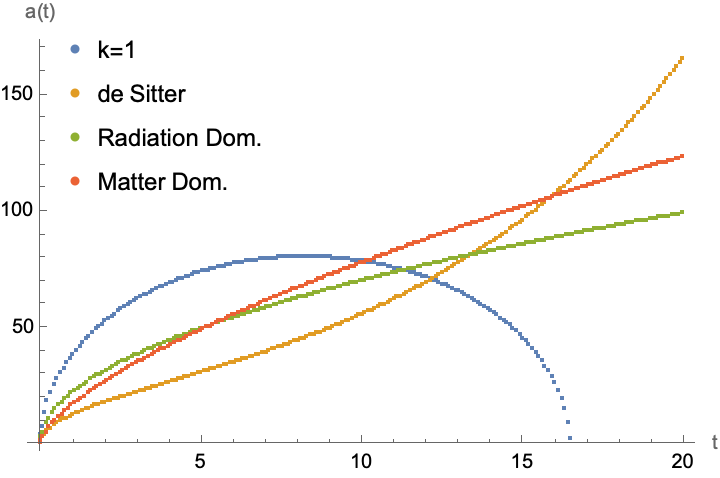}
    \caption{Representative simulated $a(t)$ for $k=1$, de Sitter, radiation dominated, and matter dominated universes. $k=-1$ omitted for neatness.}
    \label{ScaleExamples}
\end{figure}

We can see how a time slice of the simulation looks in FIG. \ref{3D}. It show a universe with particle tracks in both physical and comoving coordinates, color-coded according to time. The color-coding is also seen in the data for $a(t)$ below, so the value of the $a$ at any instance in a particle's trajectory can be read off. The data for $a$ are fitted to a power law, $a(t)\sim t^b$, from which $b$ is extracted.
As seen in FIG. \ref{ScaleExamples}, the system displays the expected scaling behavior of $a(t)$ for radiation and matter domination, $a(t)\sim t^{{2}/{3(1+w)}}$. The eras are created by initializing the matter as either HR or NR.

\begin{figure}[t]
    \includegraphics[scale=.65]{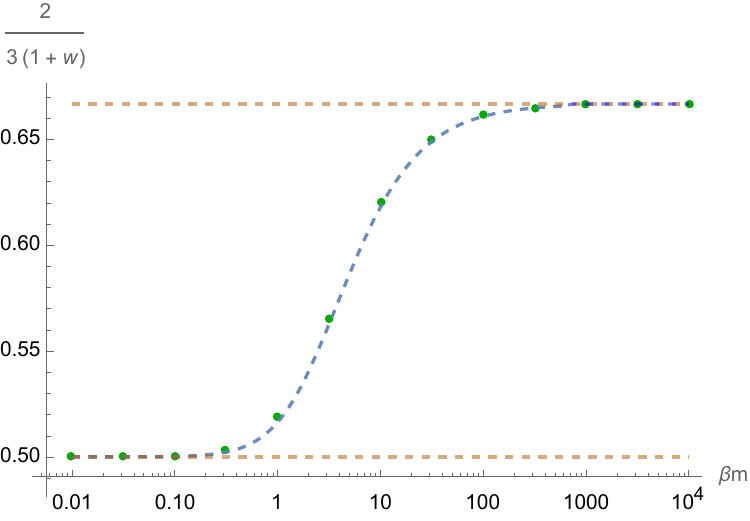}
    \caption{The simulated scaling exponent of $a(t)$ vs. the initial $\beta m$ of an $N=250$ particle non-interacting relativistic gas in equilibrium in units of $a(0)=c=G=1$. The simulation time for each is small such that $w$ is approximately constant. The horizontal lines denote the asymptotic values $1/2$ and $2/3$, corresponding to radiation and matter domination.}
    \label{Sigmoid}
\end{figure}

Additionally, we see the correct behavior for closed and de Sitter universes, corresponding to $k=1$ and $\Lambda>0$ respectively. FIG. \ref{Sigmoid} shows the scaling exponent ${2}/{3(1+w)}$ for a universe with $k=\Lambda = 0$ as a function of the initial inverse temperature and mass $\beta m$ of the matter. In this scenario, the momentum distribution of the particles of mass $m$ is initialized as a random sample of the relativistic equilibrium distribution, Eq. (\ref{RelEquilDist}). We will see in the following sections that $w$ runs with time outside of the $w=1/3$ and $w=0$ fixed points -- $a(t)\sim t^{{2}/{3(1+w)}}$ is only a solution of the Friedmann equations when $w$ is a constant, or slow varying compared to the Hubble rate. 

FIG. \ref{Sigmoid} retains validity since the time scale over which $a$ is fitted is small compared to the time scale over which $w$ runs.

$w$ decreases with expansion, corresponding to the matter losing both kinetic energy to redshift and potential energy as it spreads out. FIG. \ref{Energies} shows how the terms in Eq.~(\ref{Ham}) behave with time for a growing universe with $k=\Lambda = 0$. Generically, all the terms decrease in magnitude with time as long as the interparticle potential decreases with increasing r. $K$ tends to the sum of the rest masses and $T$ tends to $-K$, both constant, while $V$ always goes to 0 and crosses $K$ if $V(0) > K(0).$

\section{A Cosmological Gas}
The overall goal of this study is to construct theoretical models and numerics to study competing resolutions to the past hypothesis paradox. How, then, would entropy computations look? We argue in Sec. VIII that an definition of entropy suitable for non-equilibrium systems is required.  However, it will be useful to study equilibrium entropy to the extent that we can, in order to connect with arguments discussed in the introduction \cite{Padmanabhan:2008xf, Aoki, Frautschi:1981xu, 2008PhLRv...5..225L, Weber1988EntropyIA,Layzer1990-LAYCTG}.

In this section, we consider the equilibrium thermodynamics of a relativistic ideal gas in cosmology -- a ``cosmological gas." 
For the moment, we will treat $a$ as a parameter and attempt to calculate the canonical partition function for the particle terms of Eq. (\ref{Ham}). The true partition function should be microcanonical, as the energy of the universe is fixed to be 0.

The thermodynamics of a relativistic gas are known historically \cite{Juttner}, with a small resurgence in interest in the cosmological context \cite{DE_BERREDO_PEIXOTO_2005}. For a fully covariant treatment, see \cite{Remi}. de Berredo-Peixoto et. al. \cite{DE_BERREDO_PEIXOTO_2005} compute results for the ``reduced relativistic gas" in cosmology, assuming all particles have equal kinetic energies. We will relax this assumption and see that this simplified (non-covariant and collisionless) framework allows us to derive well-known freeze-out results and an explicit formula for the equation of state parameter $w$ as a function of $\beta m$, where $m$ is the mass of the gas particles.

Specializing to $k=0$, we have the following Hamiltonian for a relativistic free gas in cosmology:
\begin{equation}
 H=\sum^N_{i=1}{\sqrt{m_i^2+\frac{p_{r_i}^2}{a^2}+\frac{p_{\theta_i}^2}{a^2r_i^2}+\frac{p_{\varphi_i}^2}{a^2r^2\sin^2(\theta_i)}}}.
\end{equation}
We compute the one-particle canonical partition function after setting $k_B= \hbar = 1$,
\begin{equation}
\begin{split}
Z_1&=\int d^3x\, d^3p\, e^{-\beta \sqrt{m^2 + \frac{|\boldsymbol p|^2}{a^2}}} \\&= 4\pi V a^3 m^2 \frac{K_2(\beta m)}{\beta},
\end{split}
\end{equation}
where $K_2$ is a modified Bessel function of the 2nd kind and the integral is over comoving quantities (V is the comoving volume). This has the following HR ($\beta m \rightarrow 0$) and NR ($\beta m \rightarrow \infty$) limits, computed via asymptotic expansion:
\begin{equation}
Z^{HR}_1=8\pi V\frac{a^3}{\beta^3}
\end{equation}
and
\begin{equation}
Z^{NR}_1=Va^3 \left(2\pi m \frac{1}{\beta}\right)^{3/2}e^{-\beta m},
\end{equation}
which are the correct partition functions for HR and NR gases. The latter has a factor of $e^{-\beta m}$ due to contributions of the rest mass, which does not affect the entropy. It only affects the free and internal energies by the addition of a factor of $m$. 

The equation of state of the gas (with $Z=Z_1^N/N!$) is $PVa^3=NT$. Then, $\overbar E = - \partial \log Z/\partial \beta$, so

\begin{equation}
\overbar E= N\left(m\frac{K_3(\beta m)}{K_2(\beta m)}-\frac{1}{\beta}\right)=N\overbar E_1,
\end{equation}
where $K_3$ is also a modified Bessel functions of the 2nd kind. We write the equation of state in the form of a perfect fluid $P=w\rho,$ where $\rho=N \overbar E_1/(a^3V)$, and equate $w\rho=NT/(Va^3)$ to solve for the temperature dependence of the equation of state parameter $w$. This implies $\overbar E_1 = T/w$. $T(\overbar E_1)$ exists, but it is not possible to write explicitly. It is, however, exactly correct to write the equation of state as $PVa^3=NT(\overbar E_1),$ or equivalently $PVa^3=Nw\overbar E_1.$

We then have
\begin{equation}\label{w}
    w=\frac{1}{\beta m \frac{K_3(\beta m)}{K_2(\beta m)}-1},
\end{equation} with the limits $w\rightarrow \frac{1}{3}$ as $\beta m \rightarrow 0$, and $w\rightarrow 0$ as $\beta m \rightarrow \infty$. This function nicely fits the data in FIG. \ref{Sigmoid}, which was simulated for short enough times such that $w$ is approximately constant.
Once in thermal equilibrium, this gas is isotropic and homogeneous. It can be trivially shown that, for the collisionless case, the viscosity vanishes and the thermal conductivity vanishes to first order~\cite{Israel}. The gas can therefore be considered a perfect fluid, and the $w$ in the gas equation of state corresponds to the $w$ in the perfect fluid equation of state.

\begin{figure}
    \includegraphics[scale=.65]{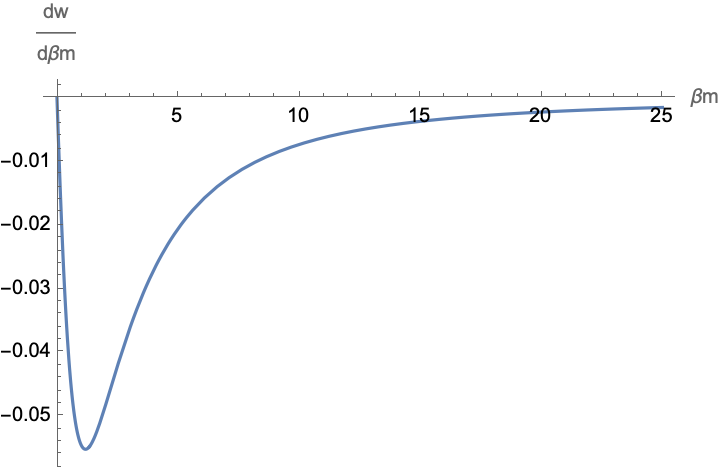}
    \caption{$\frac{dw}{d(\beta m)}$. It has fixed points at $\beta m = 0$ and $\infty$, corresponding to $w=\frac{1}{3}$ and $w=0$, respectively.}
    \label{dwdbm}
\end{figure}

The equilibrium entropy is $S=\beta(\overbar E-F)$, which we can compute as $S=\log Z +\frac{N}{w}-\log N!,$ which is
\begin{equation}
\begin{aligned}
S=N\log\, \left(4\pi V a^3 m^2 \frac{K_2(\beta m)}{\beta}\right)\\ + N\left(\beta m \frac{K_3(\beta m)}{K_2(\beta m)}-1\right)-\log N!.\\
\end{aligned}
\end{equation}

It has been proven that a relativistic gas cannot undergo adiabatic expansion, in which the entropy is conserved \cite{Abellan}. This non-conservation can also be shown numerically using the relationship  Eq.~(\ref{MiddleTempScaling}) between $\beta$ and $a$ derived in the following section. However, it has also been argued that this effect is small and not of critical importance in cosmology \cite{1970CoASP...2..121S, 1970CoASP...2..206S}, and thus likely not a large contributor to the story of entropy in an expanding universe.

\section{Studying Non-Equilibrium Statistical Mechanics}

The most critical feature in both of the arguments in the introduction is the non-equilibrium nature of thermodynamics in gravitational systems. In this section, we will show how this behavior emerges in the cosmological context via our model and how some known thermodynamic results arise from the cosmological gas.

The interplay of how fast $a$ changes, versus how fast matter interacts, is the basis for the cosmological phenomenon of \textit{freeze-out}, which describes the process of an equilibrium system becoming pseudo-equilibrium or unable to equilibrate due to expansion.  FIGS. 6, \ref{FreezeCombined} and \ref{FreezeBetas} depict this process. 

The condition defining freeze-out is
\begin{equation}\label{freeze}
    \frac{\dot a}{a} \approx \frac{N}{Va^3}
    \langle\sigma v\rangle,
\end{equation}
where $N/(Va^3)$ is the number density, $\sigma$ is the interaction cross-section, and $v$ is the comoving velocity distribution. The classical scattering cross section is defined as 
\begin{equation}
    \int \sin \theta \,d\theta\, d\phi\, \frac{b}{\sin \theta}\left|\frac{db}{d\theta}\right|,
\end{equation}
where $b$ is the impact parameter. The term $\dot a/a$ scales as $t^{-1}$ and the right hand side decays faster than $t^{-3/2}$. At early times, the Hubble rate is smaller than the interaction rate, and the freeze-out happens when these rates intersect. After this, the matter is increasingly unable to interact effectively due to the distance between the particles increasing and the speeds decreasing. In the following we consider \textit{kinetic freeze-out}, in which the elastic scattering that transfers particle momenta ceases. This is in contrast to \textit{chemical freeze-out}, which occurs at a higher temperature, when the inelastic processes that change particle species cease \cite{Du_2022}.

\begin{figure}\label{Freezegaussians}
    \includegraphics[scale=.7]{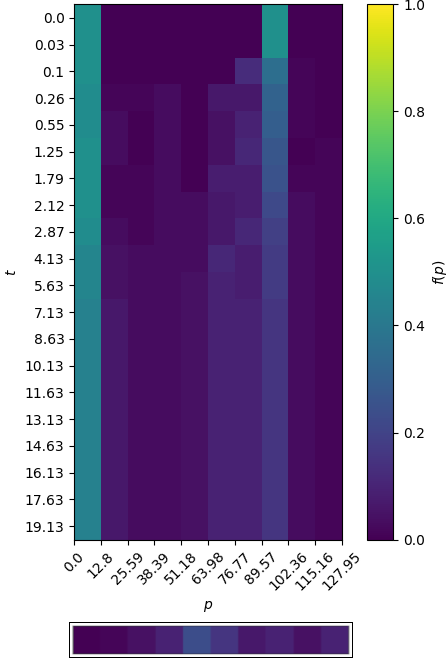}
    \caption{Momentum histogram showing the freeze-out of a single species as it starts out of equilibrium and tries to thermalize but fails. The distribution starts as 2 gaussians and attempts to reach the equilibrium distribution, which is boxed below.}
\end{figure}

A cosmological gas at equilibrium has the following 1-particle momentum distribution:
\begin{equation}\label{RelEquilDist}
\begin{split}
f\, d^3p &= \frac{e^{-\beta \sqrt{m^2 + \frac{|\boldsymbol p|^2}{a^2}}}}{Z_1/V} d^3p \\&= \frac{4\pi |\boldsymbol p|^2e^{-\beta \sqrt{m^2 + \frac{|\boldsymbol p|^2}{a^2}}}}{Z_1/V} d|\boldsymbol p|,
\end{split}
\end{equation}
and for a given $a$, there is only 1 parameter: $\beta$. 
For times when the particle scattering rate is much greater than the expansion rate, this is the momentum distribution and the equilibrium distribution.

The corresponding HR and NR distributions are

\begin{equation}\label{NRDist}
f^{NR}\,d^3p =\left(\frac{\beta }{a^2}\right)^{3/2} \frac{4\pi |\boldsymbol p|^2e^{-\beta \frac{|\boldsymbol p|^2}{2ma^2}}}{(2\pi m)^{3/2}} d|\boldsymbol p|
\end{equation}
and 
\begin{equation}\label{HRDist}
f^{HR}\,d^3p =\left(\frac{\beta}{a}\right)^3 \frac{4\pi |\boldsymbol p|^2e^{-\beta \frac{|\boldsymbol p|}{a}}}{8\pi} d|\boldsymbol p|.
\end{equation}

After freeze-out, particles the keep the same comoving momentum distribution, Eq. (\ref{RelEquilDist}) if in equilibrium. The condition for maintaining the distribution is $\frac{df}{dt}=0.$ A particle species that started in equilibrium will retain an equilibrium distribution, but will be out of equilibrium with other matter. 

We are now able to complete common lore regarding temperature scaling during expansion. We see that Eqs. (\ref{NRDist}) and (\ref{HRDist}) can be written in terms of a new variable $\beta/a^\alpha,$ where $\alpha$ is $1$ or $2$. After freeze-out, the distributions do not change, so $\beta(t)/{a(t)^\alpha}=\beta_*/{a_*^\alpha}$, where the star denotes the values at freeze-out. This means that $T= T_* a_*^\alpha/{a^\alpha}$ (here, $T$ is really a sort of ``effective" temperature: that which the distribution matches the thermal equilibrium distribution at that temperature.) In the intermediate regime, the temperature has a more complicated relationship with $a$, via
\begin{equation}\label{MiddleTempScaling}
\frac{d}{dt} \left (\frac{\beta \, e^{-\beta \sqrt{m^2 + \frac{1}{a^2}}}}{a^3 m^2 K_2(\beta m)}\right )=0.
\end{equation}

This intermediate regime is exactly where $w$ runs. As we saw, the $w$ in the HR limit $\beta m \rightarrow 0$ is exactly $1/3$ and \textit{loses} all of its $\beta$ dependence, so any change of $\beta$, like the one derived above, is ignored. The same is true for the NR limit, as $w=0$ without any $\beta$ dependence. The details of this running can be computed with Eq. (\ref{w}) and Eq. (\ref{MiddleTempScaling}). As seen in FIG. \ref{dwdbm}, $w(\beta m)$ has fixed points at $\beta m = 0$ and $\infty$, corresponding to $w=\frac{1}{3}$ and $w=0$, respectively. In between, $\frac{dw}{d(\beta m)}$ is negative, so as $\beta m$ increases, as it does monotonically with $a$, $w$ decreases to $0$. The above arguments can easily be applied to a collapsing universe, in which NR matter becomes relativistic.

We have come to the conclusion, then, that when a cosmology includes matter that goes from relativistic to non-relativistic, it is correct and possibly necessary to include a source for the Friedmann equations that scales with $a$ as $(1/a^3)\sqrt{m^2+{|\boldsymbol p|}^2/a^2}$. Following Eq. (\ref{rho}), we can see the full source for a given particle species with mass $m$ would require knowing the comoving momentum distribution of those particles, as in
\begin{equation}\label{CorrectFried}
\begin{aligned}
    H^2\sim\frac{1}{a^3}\int dp\, f(p,a) \sqrt{m^2+\frac{{|\boldsymbol p|}^2}{a^2}}\\+\,\Omega_{r,0}a^{-4}+\Omega_{m,0}a^{-3}+...
\end{aligned}
\end{equation}
where we have written the $a$ dependence of $f$ explicitly, which acts as a time.

 As pure radiation and pure matter are fixed points of the flow of $w$, $\Omega_{r}$ and $\Omega_{m}$ will be time independent in the usual writing of the equation, up to particle conversion effects. However, there is no account for the flow of $w$ into the $w=0$ regime from matter with $1/3 > w >0$. This is accounted for with the suggested term, and both the radiation and matter terms can be written as the HR and NR limits of the new term. The standard way of writing the Friedmann equations works when the sources are close to the HR and NR limits, meaning that the variation in $w$ is small, as can be seen from FIG. \ref{dwdbm}.

\begin{figure*}\label{EquilCombined}
\begin{centering}
    \includegraphics[scale=.8]{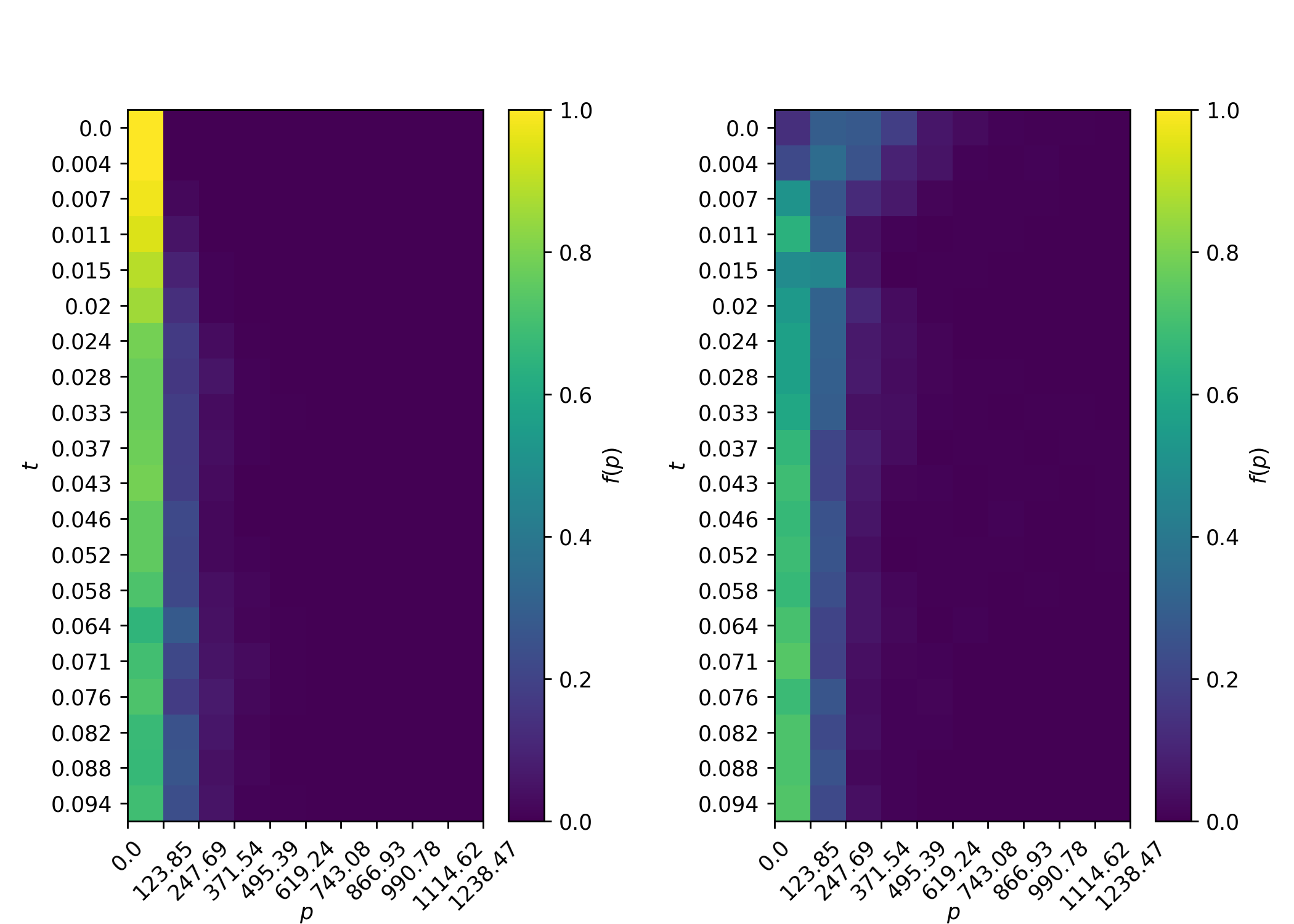}
    \caption{Time-dependent momentum histogram showing the equilibriation of two species, which appears to be around $t\sim 0.045,$ afterwhich fluctuations around the equilibrium occur. The corresponding inverse temperatures can be seen in FIG. \ref{EquilBetas}}
    \label{EquilCombined}
\end{centering}
\end{figure*}

\begin{figure}\label{EquilBetas}
    \includegraphics[scale=.65]{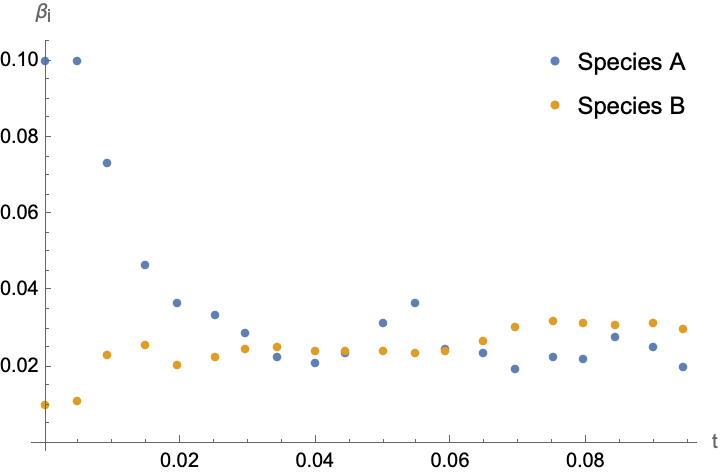}
    \caption{Time-dependent inverse temperature plot of the system in FIG. \ref{EquilCombined} Equilibrium appears to be achieved around $t\sim 0.03$, afterwhich fluctuations around the equilibrium occur.}
    \label{EquilBetas}
\end{figure}

\subsection{Numerical Results}
To study freeze-out, the Hamiltonian is modified to include an additional species of particle. The momenta are histogrammed at regular time steps and fitted with the equilibrium distribution, Eq. (\ref{RelEquilDist}), from which $\beta$ is extracted. 

The form of the potential affects the thermalization time, but qualitatively the results are independent of the interaction Hamiltonian, due to the periodicity of the simulated system. Periodicity implies that the position dependent integrals that enter the partition function will cancel explicitly with those in the equilibrium distribution, since each is finite. 

We consider a scenario analogous to the early universe: the interaction of two species, initially out of equilibrium, but each in self-equilibrium. 

FIGS. \ref{EquilCombined} and \ref{EquilBetas} show two species converge (to a temperature $T\sim (T_A+T_B)/2$)
on a much shorter timescale than that of the expansion, and FIGS. \ref{FreezeCombined} and \ref{FreezeBetas} show the converse scenario - that of freeze-out. The temperatures are drawn to each other, but we can see they start to diverge around the freeze-out time, $t_*\approx 0.4$. After this, the temperatures scale according to the relation Eq. (\ref{MiddleTempScaling}).

It should be stressed that freeze-out is a \textit{dynamical} process, with no strict cut-off between equilibrium and non-equilibrium states. Namely, there is a smooth transition between the regimes of the left-hand and right-hand sides dominating the relation Eq. (\ref{freeze}). The process is highly-nonlinear, and a complete computation to capture non-equilibrium effects would require the Boltzmann equation.
It can be seen in FIGS. \ref{EquilCombined} - \ref{FreezeBetas} that the freeze-out and equilibrium times are not strictly well-defined, following the reasoning above.

\begin{figure*}\label{FreezeCombined}
\begin{centering}
    \includegraphics[scale=.8]{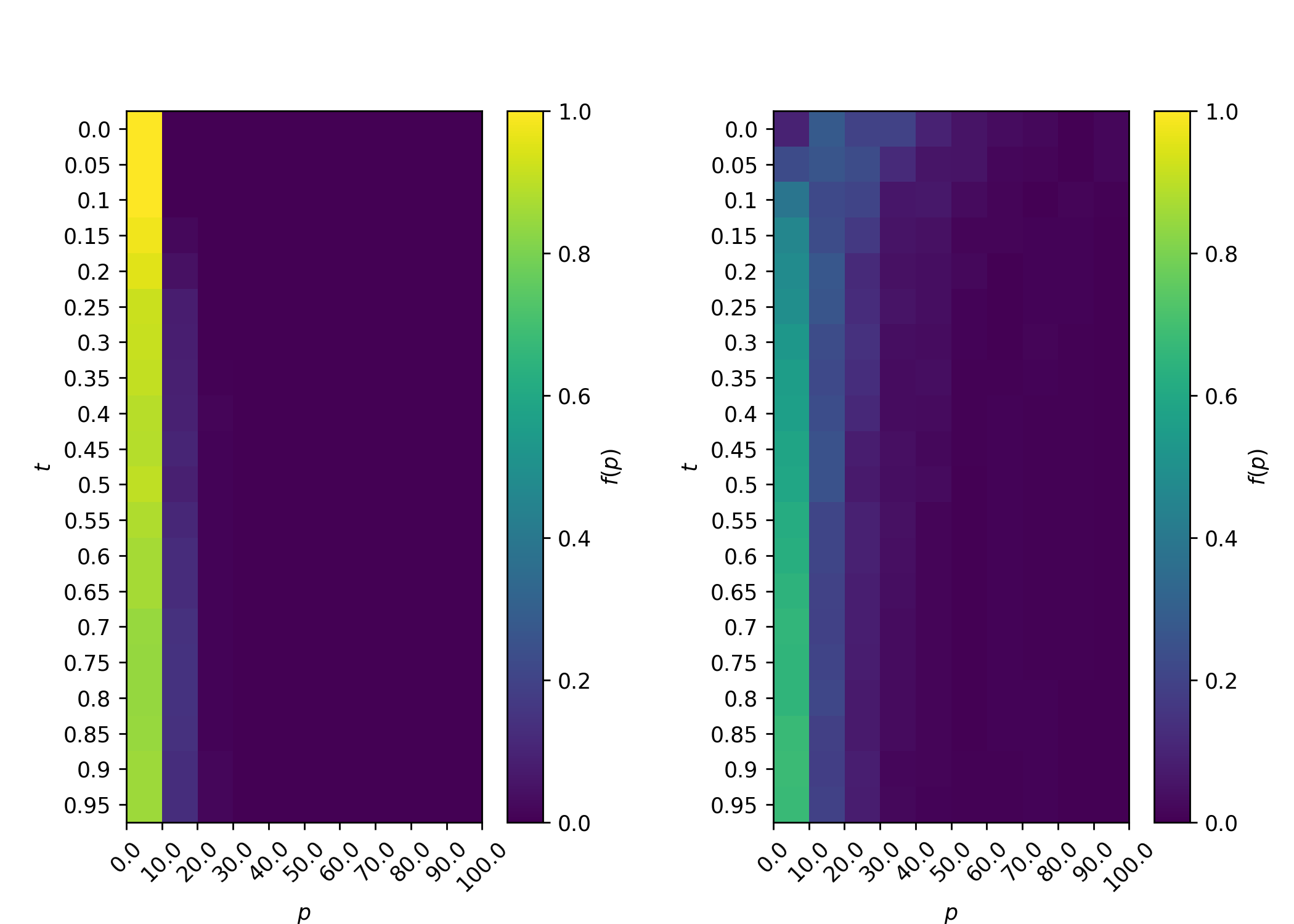}
    \caption{Time-dependent momentum histogram showing the freeze-out of two species as they starts out of equilibrium and try to thermalize but fail. Freeze-out can be seen near $t\sim 0.55$, after which the histograms cease to change. The corresponding inverse temperatures can be seen in FIG. \ref{FreezeBetas}}
    \label{FreezeCombined}
\end{centering}
\end{figure*}

\begin{figure}\label{FreezeBetas}
    \includegraphics[scale=.65]{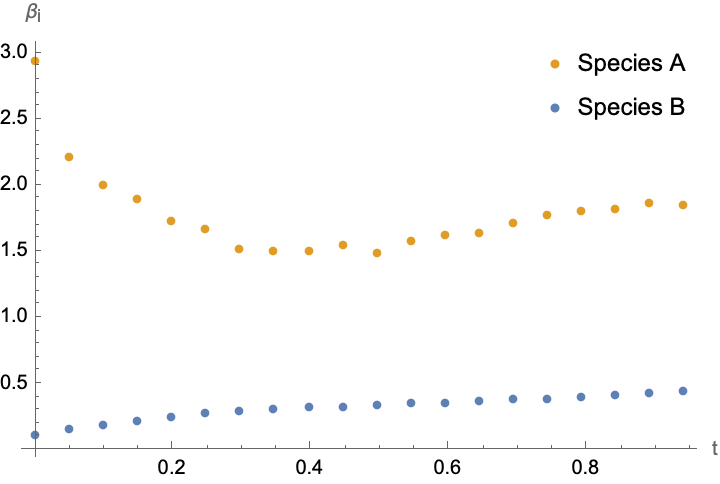}
    \caption{Time-dependent inverse temperature plot of the system in FIG. \ref{FreezeCombined} Freeze-out can be seen near $t\sim 0.40$, near when a rate of change of $\beta$ can be seen for both species.}
    \label{FreezeBetas}
\end{figure}

\section{Discussion}

The end goal of this work is a complete framework with which to compute the time-dependent entropies needed to provide numerical support for proposed solutions to the past hypothesis paradox. 

Due to the non-equilibrium nature of both cosmology and gravitational clustering, a suitable definition of entropy is needed. Coarse-grained entropies are a general class of such non-equilibrium entropies and a proper candidate would be useful for studying classical or quantum gravitational models like the one proposed.

A coarse-graining is a partitioning of the state space into sub-regions called macrostates, such that the union of all macrostates is the entirety of the state space. The exact state of the system is called the microstate, and is generally not known. The macrostates are defined according to whichever external observables are measurable. In classical mechanics, microstates are in a unique macrostate, but this is no longer true in quantum mechanics.

It is critical to note that entropy can mean many things, and it generally depends on the coarse-graining of the system. For a given state of a system, the entropy will be high with respect to some coarse-grainings and low with respect to others. It is thus required to coarse-grain in physically relevant observables if one wants to obtain physically relevant results. 

In recent years, observational entropy has emerged as a promising framework to unify and understand various entropies in both classical and quantum contexts \cite{Safranek:2020tgg,Safranek2019,Buscemi:2022wcd}. 
We propose that observational entropy would be helpful for the proposed calculation and, in general, helpful for understanding other aspects of gravitational entropy and entropies in gravitational systems, which is currently limited to Bekenstein-Hawking and holographic entropies.

\subsection{Phase Space Volumes and Measure Ambiguities}
In some cases, Observational and other coarse-grained entropies require integrating over $a$ in order to construct the necessary volume contributions. However, in most cosmological models, such integration is known to cause regularization issues. Famous examples of the these include \cite{Gibbons:1986xk, Gibbons_2008,Schiffrin_2012}. It has been shown that, for a homogenous scalar field in a FLRW universe, $a$ explicitly decouples from the rest of the Hamiltonian, and thus is is a ``gauge direction" in the phase space \cite{Sloan_2019}. Several solutions have been proposed, but few consider the problem fully solved. For recent literature, see \cite{Remmen:2013eja, Corichi_2014}.

The integral over $a$ can be avoided, however, by reducing the measure to a surface of constant Hubble rate, as in \cite{Corichi_2014}. This would produce an entropy as a function of the Hubble rate, which is monotonic for an expanding universe, and thus can be rewritten in terms of the time. This approach could be useful for future computations.

This is in accord with \cite{Rovelli:2019}, in which Rovelli argues that the scale factor is ``macroscopic" in the thermodynamic sense, and gives a definition for a macroscopic observable in terms of external interactions. We propose a similar understanding -- as the number of particle degrees of freedom increases, the solutions of particle motion stay chaotic, while $a(t)$ becomes \textit{less} so. This is a consequence of $\dot a (t)$ being only dependent on $a$ and the \textit{thermodynamic} variable $\rho$ in this limit. Similarly, $\ddot a(t)$ only depends on $a$ and the \textit{thermodynamic} variables $\rho$ and $P$.

\subsection{Other applications}
The proposed model could be used to more accurately probe the matter-radiation transition using Eq.~(\ref{CorrectFried}). Similarly, it is possible to compute freeze-out times much more accurately using this equation. All of the above work and arguments can be applied to contracting universes, or to other symmetry-reduced solutions of general relativity, like Bianchi models.

\section{Conclusion}
In this work, we have derived and shown properties of a cosmological model in the context of classical mechanics without general relativity. This model reproduces all features of cosmology relevant for the study of statistical mechanics, in addition to providing clarity to known thermodynamic results. New results on the thermodynamics of the equation of state were derived, in addition to how the equation of state evolves with time. 

The model has a phase space, on which the scale factor and its conjugate momentum are on equal footing with the matter, so that computations of entropy may be done. As cosmology is a non-equilibrium system, the phase space approach to computing entropies is necessary as non-equilibrium entropies are necessary. The model provides an opportunity to rigorously compare the entropy changes in cosmology associated with different gravitational effects, in order to make progress on the long-standing debate regarding what happens to entropy in the post-inflationary universe, once ``gravity is taken into account."

Our future work includes:
\begin{itemize}
    \item Constructing the volume form;
    \item Constructing a coarse-graining;
    \item Compute the observational or coarse-grained entropy as a function of time/Hubble rate;
    \item Do the above steps with and without clustering, implemented through the Newtonian gravitational potential $V\sim -Gm_1m_2/r$.
\end{itemize}

It has been known since early days of cosmology that ``Newtonian'' cosmology can be surprisingly accurate and provide an alternative perspective on well-known cosmological phenomena.~\cite{layzer1954significance}  Here we have shown that a slightly more sophisticated mechanical model can do even better, reproducing cosmological basics in a surprisingly accurate and elegant way. We hope that this can be a useful tool for shedding light on the past hypothesis paradox and, moreover, that observational entropy can be useful for understanding gravitational systems.

\section{Appendix}
\subsection{Model Lagrangian}
The proposed $\text{model}$ ($N=1$ for simplicity; $N\neq 1$ follows trivially) is derived from the Einstein-Hilbert and geodesic actions as follows:
\begin{equation}
\begin{split}
S&=S_{EH} + S_{Geo}\\&= \int d^4x \sqrt{-g} R  -m\int d\tau \sqrt{g_{\mu \nu}\dot x^\mu \dot x^\nu}, 
\end{split}
\end{equation}
where the dot denotes $\frac{d}{d\tau}.$ Choose coordinates such that $\tau=t$ and insert the $\text{FLRW}$ metric $ds^2=\mathcal{N}^2dt^2-a^2(t)\Big ((1-kr^2)^{-1}dr^2+r^2(d\theta^2 +\sin^2\theta \;{\varphi}^2)\Big )$.

Then,
\begin{equation}
\begin{aligned}
    S=V\int dt \,\mathcal{N}a^3 \left (\frac{6}{16\pi G} \left(\frac{\ddot a}{a\mathcal{N}^2} + \frac{{\dot a}^2}{a^2\mathcal{N}^2}+\frac{k}{a^2}\right )-2\Lambda \right) \\
    -m\int dt \sqrt{\mathcal{N}^2 - a^2 \left(\frac{|\dot{\boldsymbol r}|^2}{1-k{|\boldsymbol r|}^2}+r^2 \dot \theta ^2 + r^2 \sin ^2 \theta \dot \phi ^2\right)},
    \end{aligned}
    \end{equation}
    \vspace{-3cm}
indicating
\begin{equation}
\begin{aligned}
L=V\frac{1}{8\pi G}\left(\frac{3\ddot a a^2}{\mathcal{N}} + \frac{{3\dot a}^2a}{\mathcal{N}}+3ka\mathcal{N}-\Lambda a^3\mathcal{N}\right)\\ -m  \sqrt{\mathcal{N}^2 - a^2\left(\frac{|\dot{\boldsymbol r}|^2}{1-k{|\boldsymbol r|}^2}+r^2 \dot \theta ^2 + r^2 \sin ^2 \theta \dot \phi ^2\right)}.
\end{aligned}
\end{equation}
Now, we want to get rid of the $\ddot a$, so we realize that $\frac{d}{dt} (a^2 \dot a)= 2a{\dot a}^2 + a^2 \ddot a$. So, $a^2 \ddot a= \frac{d}{dt} (a^2 \dot a)-2a{\dot a}^2$. We can insert this and gauge away the total time derivative, which is a boundary term on the time integration, to arrive at
\begin{equation}
\begin{aligned}
    &L=V\frac{1}{8\pi G}\left(-\frac{{3{\dot a}^2}{a}}{\mathcal{N}}+{3k}{a}\mathcal{N}+\Lambda a^3\mathcal{N}\right)\\-m &\sqrt{\mathcal{N}^2 - a^2\left(\frac{|\dot{\boldsymbol r}|^2}{1-k{|\boldsymbol r|}^2}+r^2 \dot \theta ^2 + r^2 \sin ^2 \theta \dot \phi ^2\right)}.
\end{aligned}
\end{equation}
The Hamiltonian is obtained via a Legendre transformation, and in the discussion of the it and its equations of motion, we have set the comoving volume V to be 1 for simplicity.
\begin{figure}[t]
    \includegraphics[scale=1]{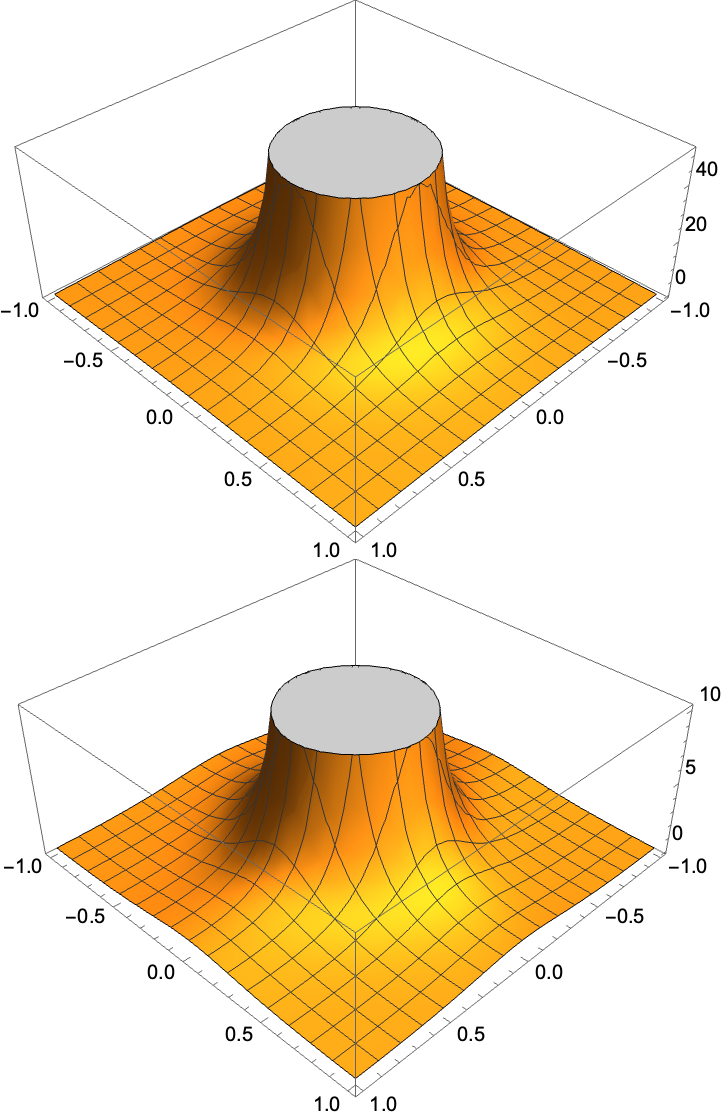}
    \caption{Comparison between the central potential (top) and simulated potential in 2 dimensions, Eq. (\ref{vij}) for $\alpha=4$ and $a=1$. Both are periodic on square intervals, but the simulated potential is also smooth at the boundary, leading to continuity of forces.}
    \label{Potentials}
\end{figure}

\subsection{Mechanical Equation of State}
To compare the thermodynamic equation of state with the mechanical equations of motion, consider the case with negligible interactions:

\begin{equation}
    \rho= \frac{1}{a^3}\left(\sum_i\sqrt{m_i^2+\frac{|\boldsymbol p_i|^2}{a^2}}\right)
\end{equation}
\begin{equation}
   P= \frac{1}{3}\left(\frac{1}{a^3}\sum_i \frac{|\boldsymbol p_i|^2/a^2}{\sqrt{m_i^2+\frac{|\boldsymbol p_i|^2}{a^2}}} \right) .
\end{equation}

In the HR limit, $\frac{|\boldsymbol p|/a}{m}\gg 1,$ so
\begin{equation}
    \rho\sim \sum \frac{1}{a^3}\frac{|\boldsymbol p|}{a}\sim \sum \frac{|\boldsymbol p|}{a^4}
\end{equation}
\begin{equation}
   P \sim \sum\frac{1}{3}\left(\frac{|\boldsymbol p|^2}{a^5}\frac{a}{|\boldsymbol p|} \right)\sim \sum \frac{1}{3}\frac{|\boldsymbol p|}{a^4}.
\end{equation}

The equation of state is $P=w \rho,$ indicating correctly that $w=1/3$. 
In the NR limit,  $\frac{|\boldsymbol p|/a}{m}\ll 1,$ so to first order, 
\begin{equation}
    \rho\sim \sum \frac{m}{a^3}
\end{equation}
and $P \ll 1$, indicating that $w \sim 0$.

\acknowledgements
We express deep thanks to David Sloan, Joshua Deutsch, Joey Schindler, and Marcell Howard for the useful discussions.

\bibliography{apssamp}

\end{document}